\documentclass{PoS}

\title{Review of recent leading-order hadronic vacuum polarization calculation}

\ShortTitle{Review of recent leading-order hadronic vacuum polarization calculation}

\author{\speaker{Zhiqing Zhang}\\
        Laboratoire de l'Acc\'el\'erateur Lin\'eaire, IN2P3/CNRS et Universit\'e Paris-Sud 11, Orsay, France\\
        E-mail: \email{zhang@lal.in2p3.fr}}

\abstract{The leading-order hadronic contribution to the muon magnetic moment anomaly $a_\mu\equiv (g_\mu-2)/2$, calculated using a dispersion integral of $e^+e^-$ annihilation data and $\tau$ data, is briefly 
reviewed. This contribution has the largest uncertainty to the predicted value of $a_\mu$, which differs from
the direct measurement by $\sim 3.6 (2.4)$ standard deviations for the $e^+e^- (\tau)$ based analysis.
Recent results and main open issues on the subject are discussed.}

\FullConference{Photon 2013,\\
		20-24 May 2013\\
		Paris, France }

\begin{document}

\section{Introduction}
\label{sec:intro}

The Standard Model (SM) has been extremely successful. The only missing particle of the SM, the Higgs boson, may have been discovered at the LHC~\cite{atlas,cms} and corroborated recently with property measurements~\cite{atlas1,cms1}. All SM predictions have been tested often to an extraordinary precision and no sign of new physics has been found with few exceptions. One such exception is the well known muon $g-2$ anomaly, $a_\mu$. The status, as it was given in the PDG 2012~\cite{pdg12} (see also~\cite{knecht}), is about 3.6 (2.4) standard deviations between the direct measurement dominated by the E821 experiment at BNL~\cite{bnl06} and the corresponding $e^+e^- (\tau)$ based SM predictions.

The SM prediction $a^{\rm SM}_\mu$ is usually decomposed into three parts
\begin{equation}
a_\mu^{\rm SM}=a^{\rm QED}_\mu+a^{\rm weak}_\mu+a^{\rm had}_\mu\,,
\end{equation}
corresponding to QED, weak and hadronic loop contributions, respectively. The dominant QED contribution includes all photonic and leptonic $(e, \mu, \tau)$ loops with the lowest-order being the classic $\alpha/2\pi$ Schwinger contribution. It has been computed recently through 5 loops and has the following numerical value~\cite{qed}:
\begin{equation}
a_\mu^{\rm QED}=(11\,658\,471.8951\pm 0.0080)\times 10^{-10}\,. 
\end{equation}

The weak part includes loop contributions involving heavy $W^\pm$, $Z$ and Higgs particles. It is suppressed by at least a factor $\alpha/\pi\cdot m^2_\mu/M^2_W\simeq 4\times 10^{-9}$. The numerical value accounting for the dominant 1- and 2-loop contributions~\cite{weak,weak1} is
\begin{equation}
a_\mu^{\rm weak}=(15.36\pm0.10)\times 10^{-10}\,,
\end{equation}
where the uncertainty stems mainly from quark triangle loops (the uncertainty due to that of the Higgs mass after its discovery, taken to be 1.5\,GeV, becomes now negligible~\cite{weak1}).

The hadronic part involving quark and gluon loop contributions may be further decomposed into leading-order (LO), higher-order (HO) and light-by-light (LBL) scattering contributions $a_\mu^{\rm had}=a_\mu^{\rm had,\, LO}+a_\mu^{\rm had,\, HO}+a_\mu^{\rm had,\, LBL}$. At present, the LO contribution cannot reliably be calculated from perturbative QCD (pQCD) and is determined instead by a dispersion relation~\cite{dispersion}
\begin{equation}
a_\mu^{\rm had,\, LO}=\frac{1}{3}\left(\frac{\alpha}{\pi}\right)^2\int^\infty_{m^2_{\pi^0\gamma}}ds\frac{K(s)}{s}R^{(0)}(s)\,,
\end{equation}
where $R^{(0)}(s)$ represents  the ratio of the bare cross sections of $e^+e^-$ annihilation into hadrons to the point-like muon-pair cross section and  $K(s)\sim 1/s$ is a QED kernel function~\cite{kqed} and gives a strong weight to low-energy part of the integrand. The precision of $a_\mu^{\rm had,\, LO}$ depends thus on that of the $e^+e^-$ annihilation data in particular that of $\rho(770)\to \pi^+\pi^-$ and it has the largest uncertainty to $a_\mu^{\rm SM}$ and this is why most of the effort from both experimental and theoretical sides went into its improved determination over the last 20 years or so.

In the following, we shall briefly describe the new development including an update of the $\tau$ spectral functions from ALEPH~\cite{aleph13} and discuss a few open issues on the subject.

\section{New development}
Three new results\footnote{Since the workshop, a few other new results have appeared. One example is the cross section measurement of $e^+e^-\to K^+K^-(\gamma)$ with the initial-state radiation method from BaBar~\cite{kk13}, which contributes to $a_\mu$ with $22.93\pm 0.18_{\rm stat}\pm 0.22_{\rm syst}$ in the energy range from threshold to 1.8\,GeV, to be compared with our previous evaluation $21.63\pm 0.27_{\rm stat}\pm 0.68_{\rm syst}$ based on other data~\cite{dhmz10}.} are discussed in this review:
\begin{enumerate}
\item New $e^+e^-\to \pi^+\pi^-$ cross section measurement from KLOE based on its 2002 data~\cite{kloenew}. Previously, the cross sections were measured based on the integrated luminosity obtained by counting Bhabha scattering events and using the corresponding QED cross section prediction~\cite{kloeold}. The new measurement is normalized to that of $e^+e^-\to \mu^+\mu^-(\gamma)$ as it was performed by BaBar~\cite{babar}; 
\item New cross section measurements from BaBar in multi-hadronic channels $e^+e^-\to 2\pi^+2\pi^-$, $K^+K^-\pi^+\pi^-$, $K^+K^-2\pi^0$~\cite{babar1};
\item Updated ALEPH non-strange spectral functions from hadronic $\tau$ decays~\cite{aleph13}.
\end{enumerate}

The new KLOE measurement has the advantage over the old one in that the systematic uncertainty (in particular the theoretical uncertainty) is substantially reduced due to cancellation in the $\pi\pi\gamma$ to $\mu\mu\gamma$ ratio measurement. Indeed, in terms of $a_\mu [\pi\pi]$ in the energy range between 0.592 and 0.975\,GeV, the new measurement contributes $385.1\pm 1.1_{\rm stat}\pm 2.6_{\rm exp}\pm 0.8_{\rm th}$ to be compared with the old value of $387.2\pm 0.5_{\rm stat}\pm 2.4_{\rm exp}\pm 2.3_{\rm th}$ (if not otherwise stated, these and the following values are given in units of $10^{-10}$).

\begin{figure}[htb]
\begin{center}
\vspace{-2mm}
\includegraphics[width=.475\textwidth]{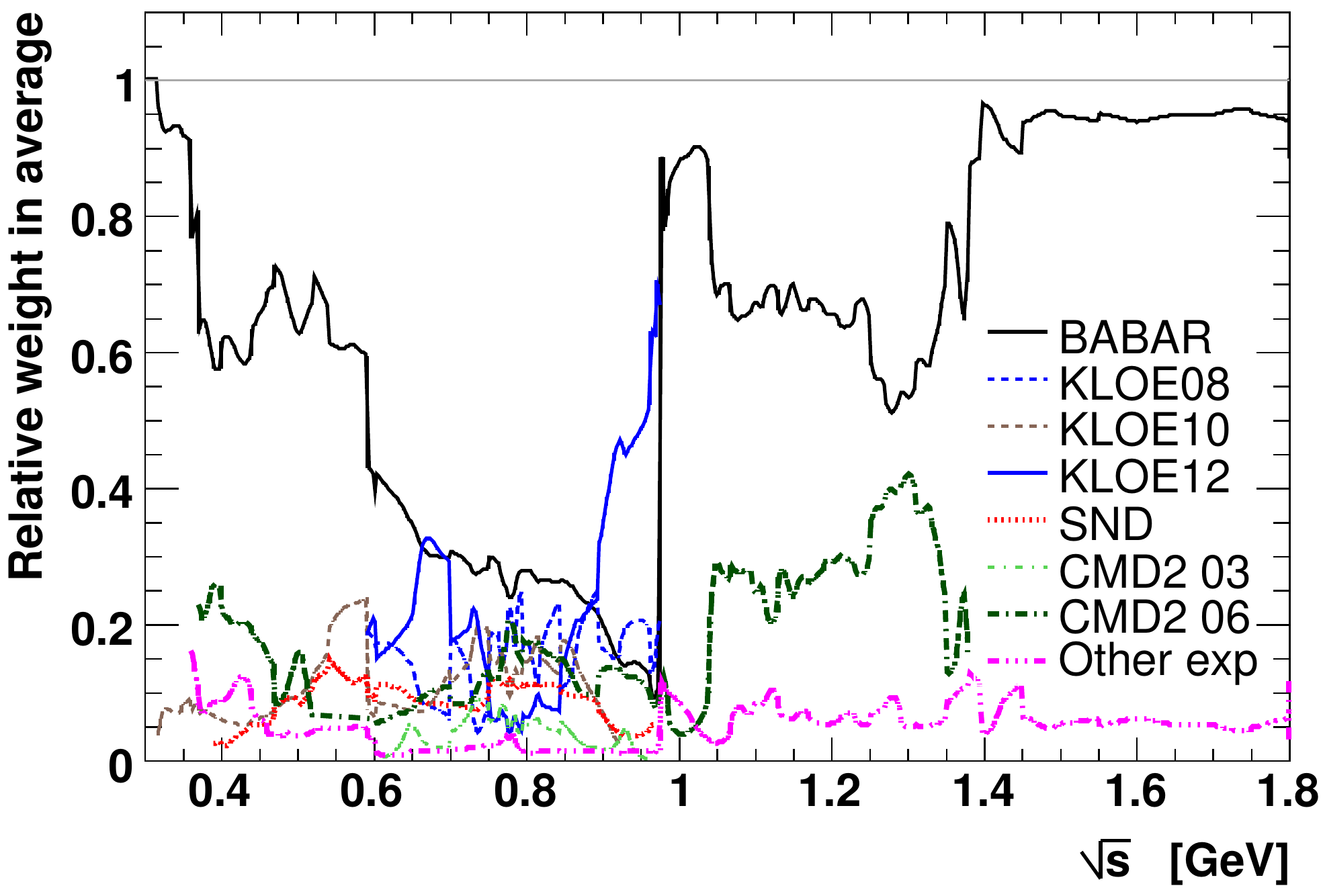}
\includegraphics[width=.475\textwidth]{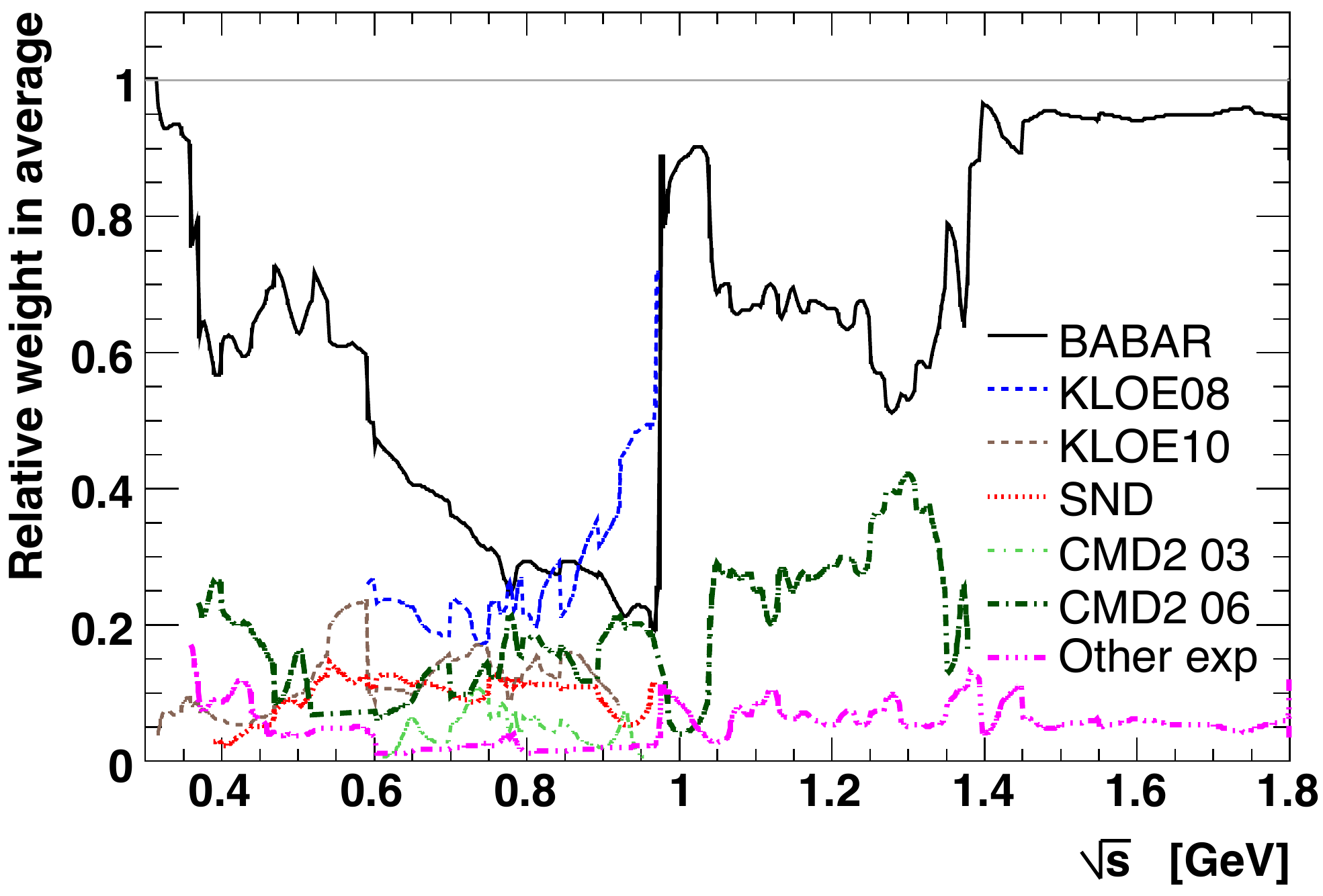}
\end{center}\vspace{-7mm}\caption{Relative local averaging weight per experiment versus centre-of-mass energy in $e^+e^-\to \pi^+\pi^-$ either including (left) or excluding (right) the new KLOE measurement (KLOE\,12). Other measurements are referenced in~\cite{dhmz10}.}
\label{fig:weight}
\end{figure}

Including the new measurement (KLOE12) in our data combination based on HVPTools~\cite{hvptools} and taking into account the correlation between this measurement with other data sets to our best knowledge\footnote{The statistical correlations with the KLOE08 measurement (from the common $2\pi$ sample) are not available and were not taken into account.}, we obtain $a_\mu[\pi\pi]=506.56\pm 2.61_{\rm stat}\pm 2.38_{\rm syst}$ in the energy range of $0.3-1.8$\,GeV to be compared with the corresponding value $507.24\pm 2.87_{\rm stat}\pm 2.56_{\rm syst}$ excluding this measurement. The relative impact of the measurement in the combination may be seen in Fig.~\ref{fig:weight}.

The second set of new results concerns multi-hadronic channels $e^+e^-\to 2\pi^+2\pi^-, K^+K^-\pi^+\pi^-$ and $K^+K^-2\pi^0$. For the first channel, using the final measurement from BaBar~\cite{babar1} together with other available data~\cite{dhmz10}, we obtain $a_\mu[2\pi^+2\pi^-]=13.64\pm 0.03_{\rm stat}\pm 0.36_{\rm syst}$ to be compared with our previous evaluation of $13.35\pm 0.10_{\rm stat}\pm 0.52_{\rm syst}$ in which a preliminary version of the BaBar measurement was used~\cite{dhmz10}. The new BaBar measurement thus improves both the statistical and systematic uncertainties of the channel. BaBar has also updated their cross section measurements for channels $K^+K^-\pi^+\pi^-$ and $K^+K^-2\pi^0$ using about twice of the previously used data sample, the corresponding $a_\mu$ are $0.35\pm 0.02$ and $0.041\pm 0.007$, respectively, to be compared with the old values of $0.35\pm 0.03$ and $0.064\pm 0.012$.

The $\pi^+\pi^-$ channel used to be limited in precision, so it was proposed in~\cite{adh98} to transform the corresponding tau spectral function through an isospin rotation to the $e^+e^-$ cross section by $\sigma^{l=1}\left(e^+e^-\to \pi^-\pi^-\right)=4\pi\alpha^2/s\cdot v(\tau^-\to \pi^-\pi^0\nu_\tau)$ and to provide an independent evaluation after accounting for all isospin breaking effects~\cite{detal09}.
Similar transformations can be made for four-pion channels. Recently, updated ALEPH non-strange spectra functions from hadronic $\tau$ decays~\cite{aleph13} have been made public~\cite{alephsf13}. Compared to the 2005 ALEPH publication~\cite{aleph05}, the main improvement results from the use of a new method~\cite{ids} to unfold the measured mass spectra from detector effects. This procedure also corrects a previous problem\footnote{We thank D.~Boito for bringing this issue to our attention~\cite{boito11}.}  in the treatment of the statistical correlations between the unfolded mass bins.

\begin{figure}[htb]
\begin{center}
\vspace{-2mm}
\includegraphics[width=.475\textwidth]{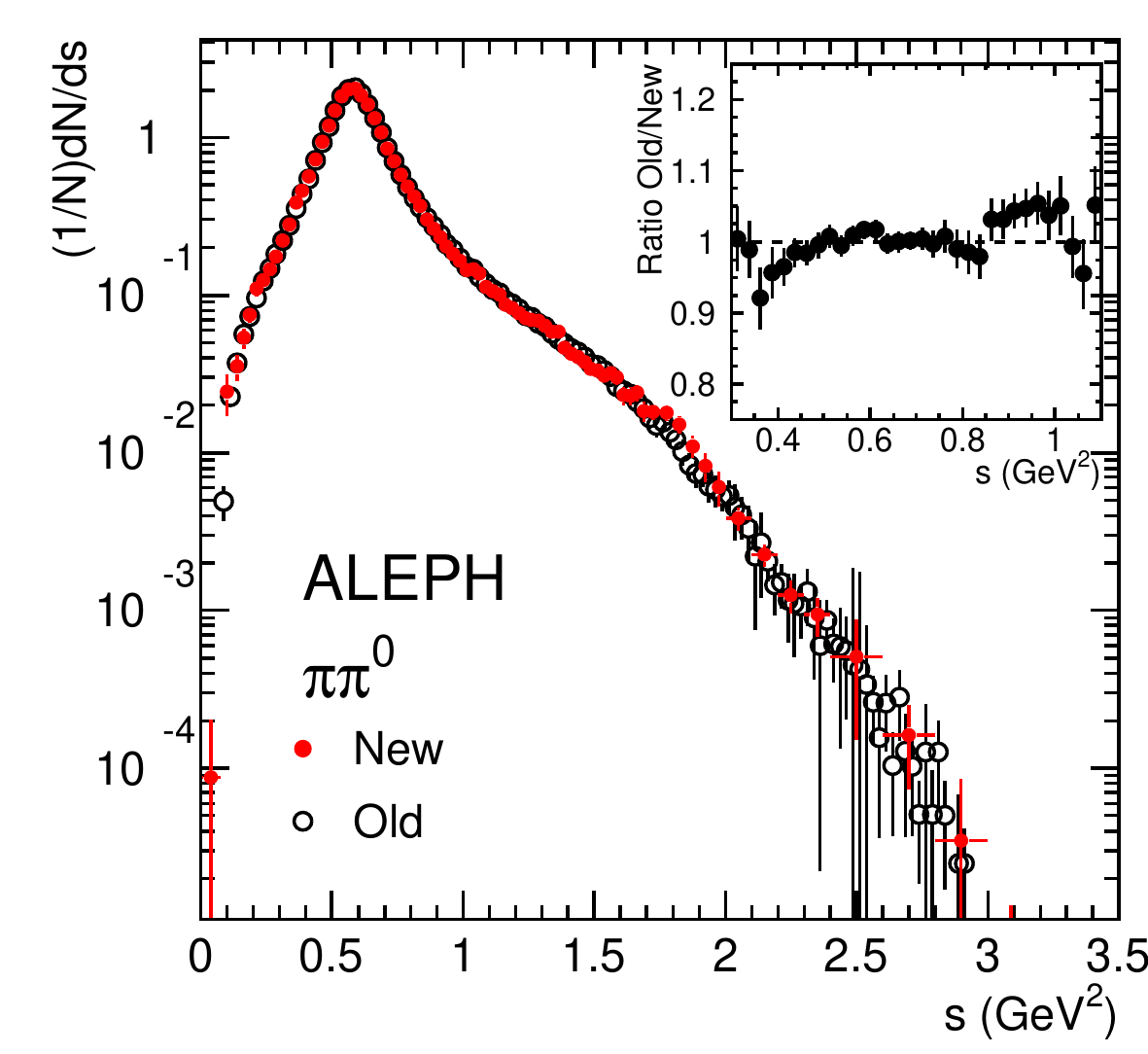}\\
\includegraphics[width=.475\textwidth]{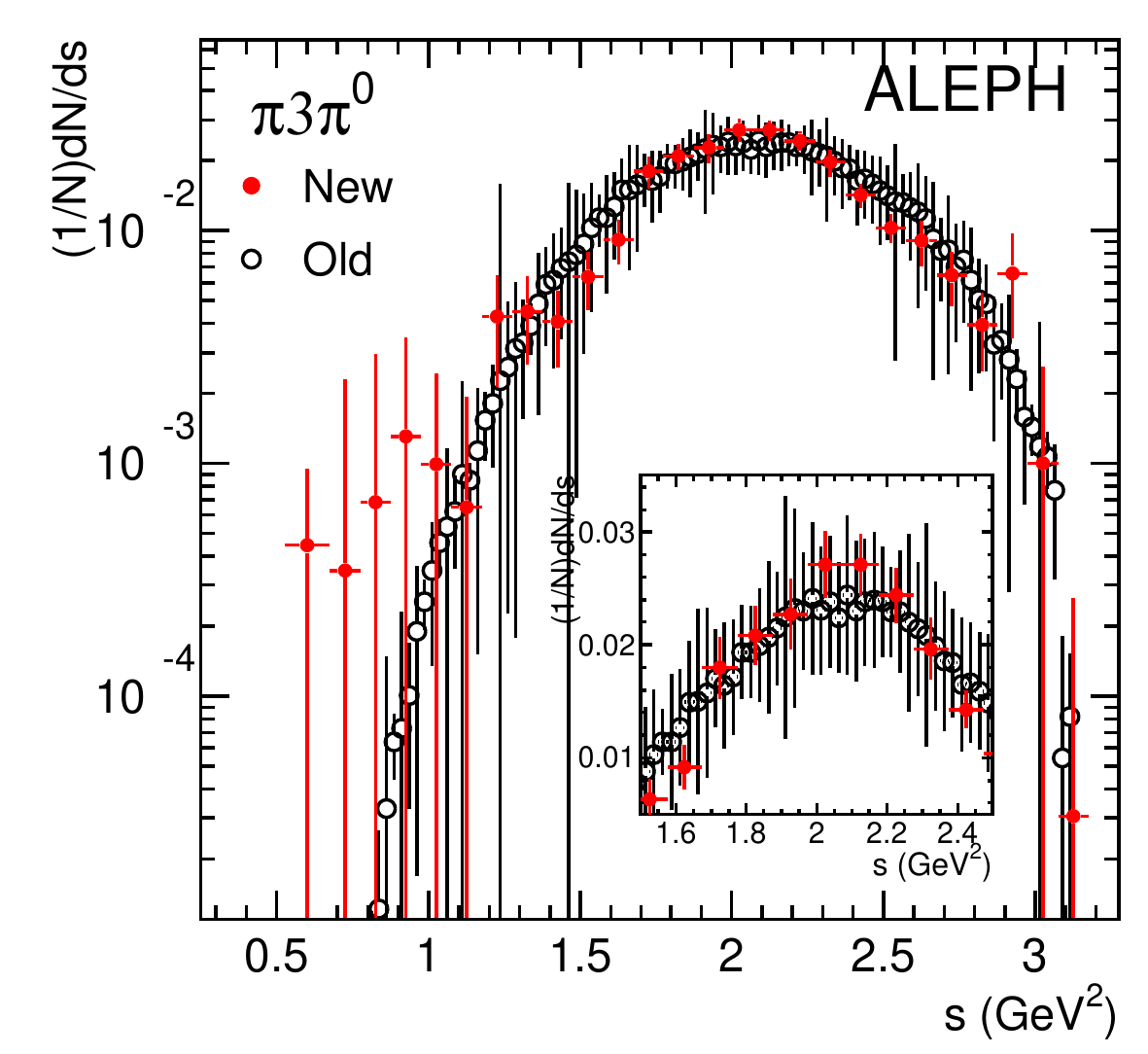}
\includegraphics[width=.475\textwidth]{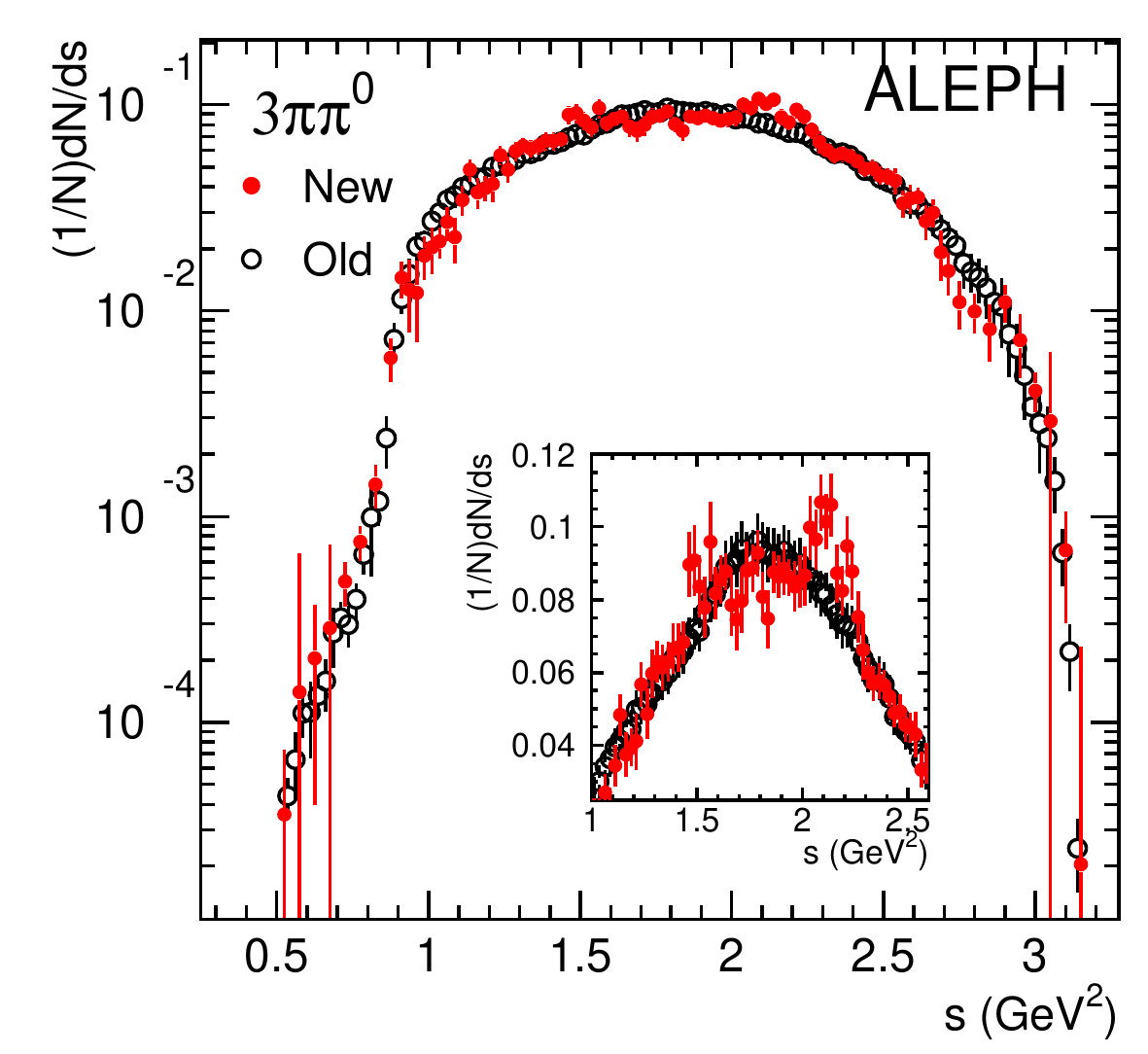}
\end{center}\vspace{-7mm}\caption{Comparison of the new unfolded spectral functions (red full circles) with those obtained in Ref.~\cite{aleph05} (black open circles). The error bars shown include statistical and all systematic uncertainties. The inset shows either the old-to-new ratios for $\pi\pi^0$ (where the error bars are those of the newly unfolded spectra) or a zoom to the peak region for the other channels.}
\label{fig:compsf}
\end{figure}

A comparison of the new unfolded mass spectra for $\pi\pi^0$, $\pi 3\pi^0$ and $3\pi \pi^0$ channels with the previous ones~\cite{aleph05} is shown in Fig.~\ref{fig:compsf}. Reasonable agreement is found everywhere except for differences at the few percent level in the $\pi\pi^0$ mode near threshold and in the $0.8-1.0\,{\rm GeV}^2$ region.
The new ALEPH $\pi\pi^0$ spectra function has been combined with those from Belle~\cite{belle}, CLEO~\cite{cleo} and OPAL~\cite{opal} for calculating $a_\mu[\pi\pi]$ with the values of $9.82\pm 0.13_{\rm exp}\pm 0.04_{\rm Br}\pm 0.07_{\rm IB}$ for the energy range between threshold and 0.36\,GeV and $506.4\pm 1.9_{\rm exp}\pm 2.2_{\rm Br}\pm 1.9_{\rm IB}$ for the energy range between 0.36 and 1.8\,GeV, where the first errors are due to the shape uncertainties of the mass spectra, which also include very small contributions from the $\tau$ mass and $|V_{ud}|$ uncertainties, the second errors originate from $B_{\pi\pi^0}$ and $B_e$, and the third errors are due to the isospin-breaking corrections, which are partially anti-correlated between the two energy ranges. These values are to be compared with the old results~\cite{detal09} of $9.76\pm 0.14_{\rm exp}\pm 0.04_{\rm Br}\pm 0.07_{\rm IB}$ and $505.5\pm 2.0_{\rm exp}\pm 2.2_{\rm Br}\pm 1.9_{\rm IB}$.

The new results for $2\pi2\pi^0$ and $4\pi$ based on linear combinations of $\tau^-\to\pi^-3\pi^0\nu_\tau$ and $\tau^-\to 2\pi^-\pi^+\pi^0\nu_\tau$, evaluated up to 1.5\,GeV, are $14.70\pm 0.28_{\rm exp}\pm 1.01_{\rm Br}\pm 0.40_{\rm IB}$ and $7.07\pm 0.41_{\rm exp}\pm 0.48_{\rm Br}\pm 0.35_{\rm IB}$, respectively, to be compared to the previous results $14.89\pm 1.22_{\rm exp}\pm 1.02_{\rm Br}\pm 0.40_{\rm IB}$ and $6.31\pm 1.32_{\rm exp}\pm 0.42_{\rm Br}\pm 0.35_{\rm IB}$. The large difference in the experimental uncertainties is not a consequence of the new unfolding but stems from a technical problem in the previous estimate of shape systematic uncertainties leading to artificially larger values.

The updated spectral functions have also been used to repeat the other analyses of~\cite{aleph05}: a phenomenological fit to the $\pi\pi^0$ mass spectrum and a QCD analysis using the vector, axial-vector and total non-strange spectral functions. The results are found in agreement with published one based on the previous set of spectral functions.

Overall the first two sets of the new results have, however, only a small impact on our previously published evaluation~\cite{dhmz10}
\begin{equation}
a_\mu^{\rm had,\, LO}=692.3\pm 1.4\pm 3.1\pm 2.4\pm 0.2\pm 0.3\label{eq:amuee}
\end{equation}
and therefore they will be included in a future reevaluation. In Eq.\,(\ref{eq:amuee}) the first error is statistical, the second channel-specific systematic, the third common systematic, correlated between at least two exclusive channels, and fourth and fifth errors stand for the narrow resonance and QCD uncertainties, respectively. 

On the other hand, the new ALEPH results do have a sizable impact on our previous evaluation. Combining the new results with those from other experiments both for tau and $e^+e^-$ data, we get
\begin{equation}
a_\mu^{\rm had,\ LO}[\tau]=703.0\pm 3.1\pm 1.9\pm 2.4\pm 0.2\pm 0.3
\end{equation}
where the first error is $\tau$ experimental, the second the uncertainty of isospin-breaking corrections~\cite{detal09}, the third $e^+e^-$ experimental, and the last two the narrow resonance and QCD uncertainties. This result is to be compared with the previous one $701.5\pm 3.5\pm 1.9\pm 2.4\pm 0.2\pm 0.3$~\cite{dhmz10}. The $2\pi$ and $4\pi$ channels account for about 78\% of the LO hadronic contribution, the rest is taken from the $e^+e^-$ channels and pQCD calculations. 

\begin{figure}[htb]
\begin{center}
\vspace{-2mm}
\includegraphics[width=.475\textwidth]{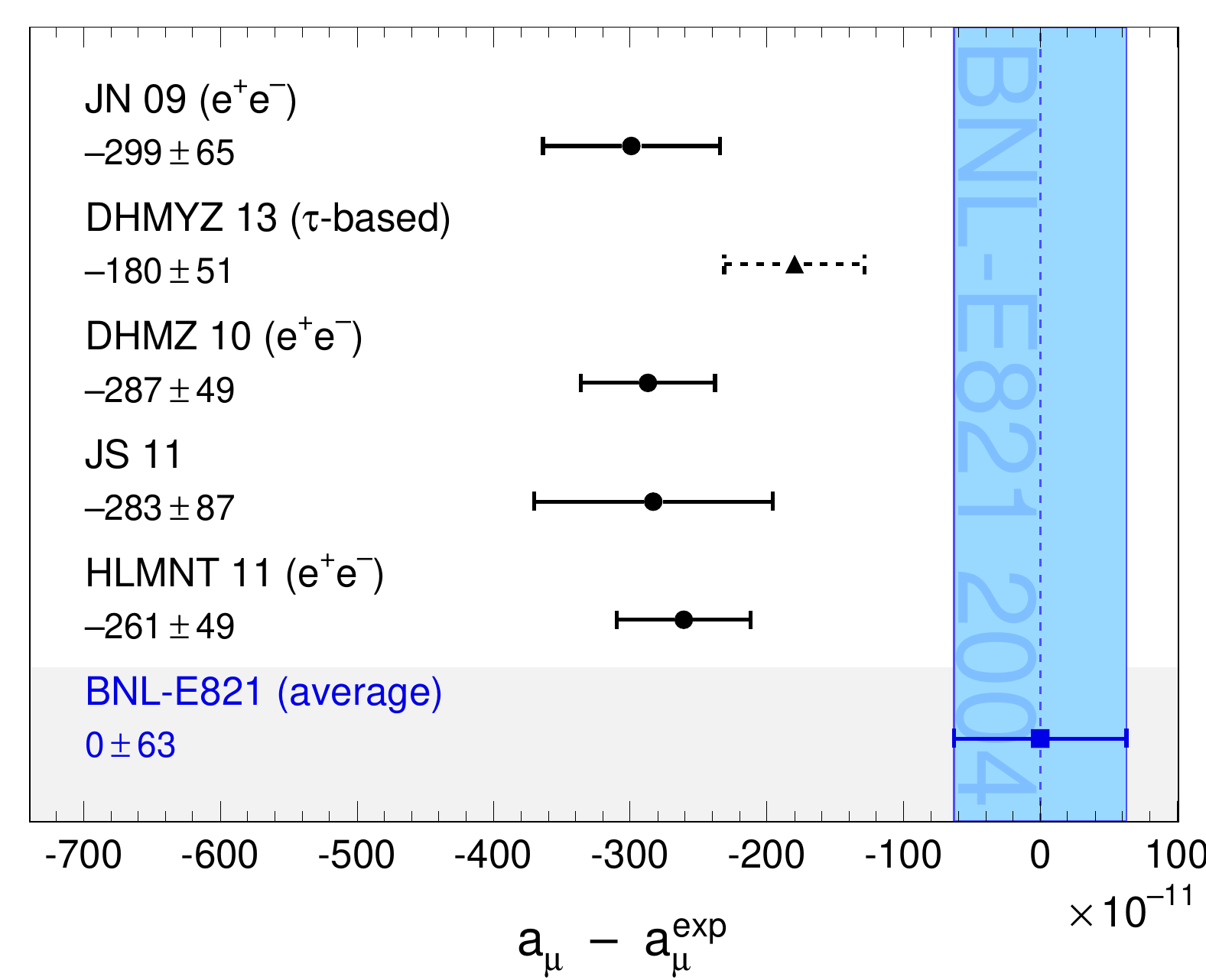}
\end{center}\vspace{-7mm}\caption{Compilation of recent results for $a_\mu^{\rm SM}$ (in units of $10^{-11}$), subtracted by the central value of the experimental average. The shaded vertical band indicates the experimental error.}
\label{fig:amures}
\end{figure}

Adding to these results the contributions from $a_\mu^{\rm had,\, HO}=-9.84\pm 0.07$~\cite{hlmnt11}, computed using a similar dispersion relation approach, $a_\mu^{\rm had,\, LBL}=10.5\pm 2.6$~\cite{lbl}, estimated from theoretical model calculations, as well as $a_\mu^{\rm QED}$ and $a_\mu^{\rm weak}$, one gets 
\begin{eqnarray}
a_\mu^{\rm SM}[e^+e^-]\!\!&=&\!\!11\,659\,180.2\pm 4.9_{\rm tot}\,,\label{eq:ee}\\
a_\mu^{\rm SM}[\tau]\!\!&=&\!\!11\,659\,190.9\pm 5.1_{\rm tot}\,.\label{eq:tau}
\end{eqnarray}
The $e^+e^-$ ($\tau$) based prediction deviates from the direct experimental average~\cite{bnl06} of
\begin{equation}
a_\mu^{\rm exp}=11\,659\,208.9\pm 5.4_{\rm stat}\pm 3.3_{\rm syst}
\end{equation}
by $28.7\pm 8.0$ $(18.0\pm 8.1)$, i.e.\ $3.6\,\sigma$ $(2.2\,\sigma)$. A compilation of recent $a_\mu^{\rm SM}$ predictions in comparison with the experimental average of direct measurements is shown in Fig.\,\ref{fig:amures}. 

\section{Open issues}

In this section, we discuss a few open issues which need to be resolved or understood before more precise direct measurements will become available.

In Fig.~\ref{fig:amures}, all recent $e^+e^-$ based predictions are consistent since the input $e^+e^-$ data sets used are largely identical. Take the comparison between DHMZ\,10~\cite{dhmz10} and HLMNT\,11~\cite{hlmnt11} as an example, they differ mainly in the data combination and error treatment. This is reflected in Table~\ref{tab:comp} (extracted from Table~4 in~\cite{hlmnt11}). The difference is comparable to or larger than one of the quoted errors. In addition, the quoted errors are quite different. It is desirable that these differences can be understood and reduced in the future. 
\begin{table}[htb]
\begin{center}
\begin{tabular}{crrr}
Channel & HLMNT\,11 & DHMZ\,10 & diff. \\\hline
$K^+K^-$ & $22.09\pm 0.46$ & $21.63\pm 0.73$ & 0.46 \\
$\pi^+\pi^-$ & $505.65\pm 3.09$ & $507.80\pm 2.84$ & $-2.15$ \\
$\pi^+\pi^-\pi^0$ & $47.38\pm 0.99$ & $46.00\pm1.48$ & 1.38\\\hline
\end{tabular}
\caption{Comparison for hadronic contributions to $a_\mu$ in the energy range from 0.305 to 1.8\,GeV from three $K^+K^-$, $\pi^+\pi^-$ and $\pi^+\pi^-\pi^0$ channels, extracted from Table~4 in~\cite{hlmnt11}.}
\label{tab:comp}
\end{center}
\end{table}

\begin{figure}[htb]
\begin{center}
\includegraphics[width=.475\textwidth]{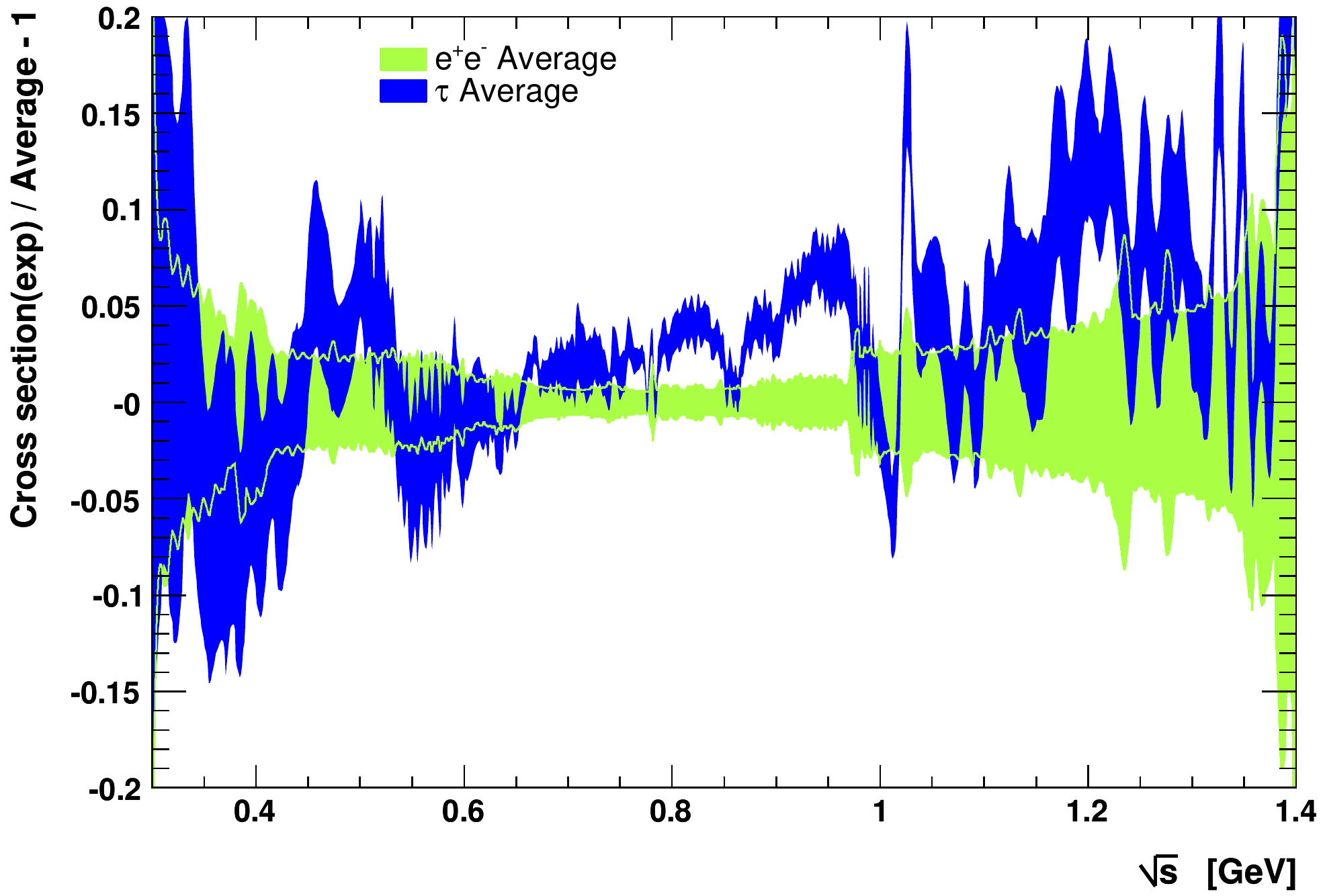}
\end{center}\vspace{-7mm}\caption{Relative shape comparison between ALEPH-Belle-CLEO-OPAL combined $\tau$ (dark shaded) and $e^+e^-$ spectral function (light shaded).}
\label{fig:ee_tau}
\end{figure}

The difference of $10.7\pm 4.9$, i.e.\ $2.2\,\sigma$, between the $\tau$ and $e^+e^-$ based predictions shown in Eqs.\,(\ref{eq:ee}) and (\ref{eq:tau}),  is another open issue. Jegerlehner and Szafron claim that the difference can be explained by the $\rho^0-\gamma$ mixing missing in the $\tau$ data~\cite{js11}. It remains to be checked whether this is the real explanation or there are experimental issues related to the $e^+e^-$ and $\tau$ measurements. Indeed, the $e^+e^-$ and $\tau$ difference can be seen from the relative shape comparison in the energy range between 0.3 and 1.4\,GeV in Fig.\,\ref{fig:ee_tau}.

\begin{figure}[htb]
\begin{center}
\includegraphics[width=.475\textwidth]{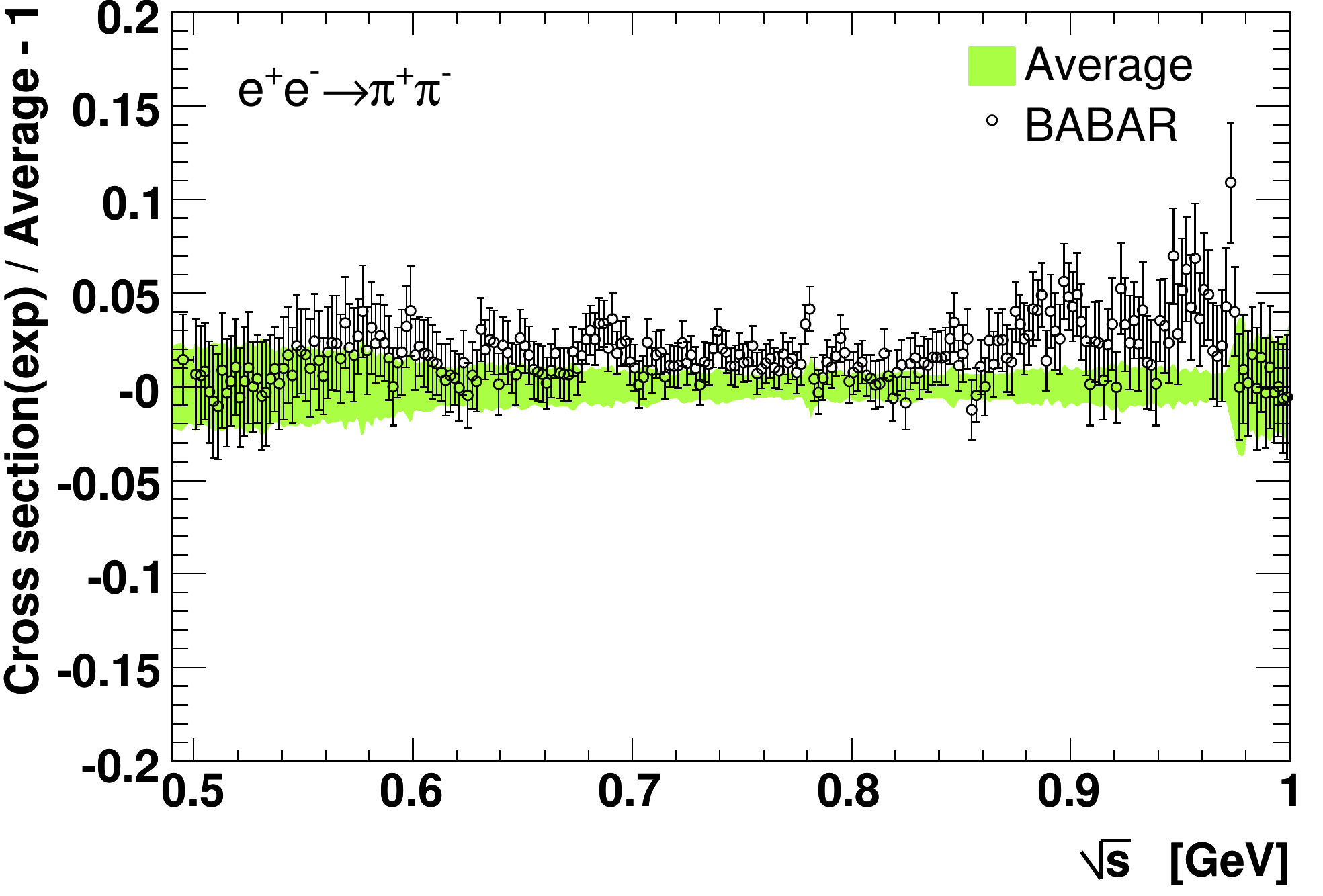}
\includegraphics[width=.475\textwidth]{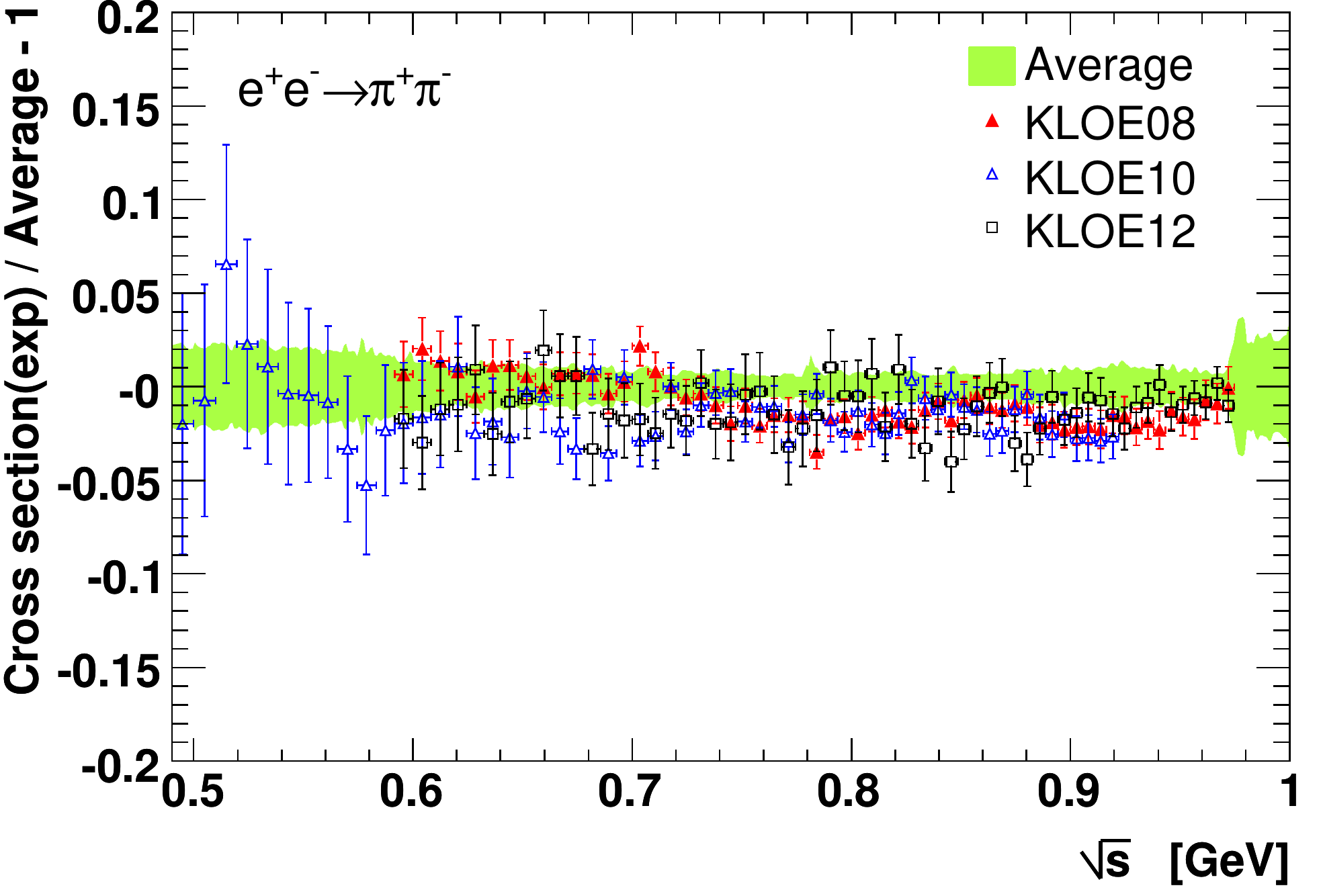}
\end{center}\vspace{-7mm}\caption{Comparison between individual $e^+e^-\to \pi^+\pi^-$ cross section measurements from BABAR (top) and KLOE (bottom) and the HVPTools average.}
\label{fig:babar_kloe}
\end{figure}

The other related issue is the different shape between BABAR and KLOE $\pi^+\pi^-$ cross section data (Fig.\,\ref{fig:babar_kloe}). This difference, leading to an amplified uncertainty in the combination following the PDG prescription, prevents further error reduction. Among the three data sets from KLOE, KLOE\,08 and KLOE\,10 were normalized using luminosity from Bhabha, whereas KLOE\,12 was performed involving the ratio of pion-to-muon pairs as BABAR did. 

\begin{figure}[htb]
\begin{center}
\includegraphics[width=.475\textwidth]{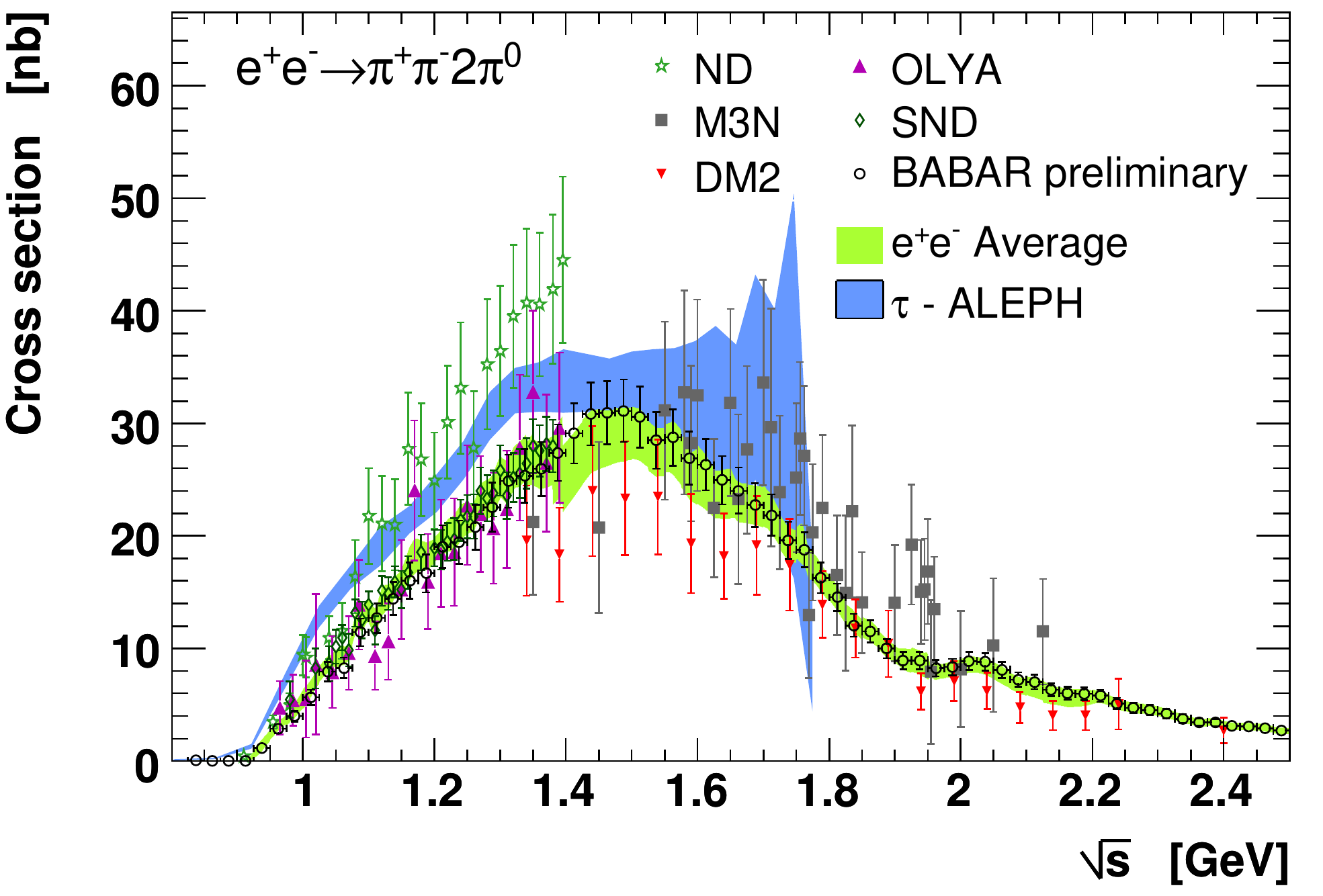}
\end{center}\vspace{-7mm}\caption{Cross section of $e^+e^-\to 2\pi^+2\pi^0$ versus centre-of-mass energy. The shaded green (blue) band gives the HVPTools average for $e^+e^-$ ($\tau$) data.}
\label{fig:2pi2pi0_ee_tau}
\end{figure}

Another problematic channel concerns $e^+e^-\to \pi^+\pi^-2\pi^0$ (Fig.\,\ref{fig:2pi2pi0_ee_tau}). There is a large scattering between measurements from different experiments, in particular between ND and other experiments. In addition when comparing the $e^+e^-$ average with the $\tau$ average, there is a significant difference in normalization. This discrepancy deserves further studies and clarification.

\section{Summary and perspectives}
The deviation of about 3.6\,$\sigma$ between the direct measurement and the $e^+e^-$ based SM predictions on $a_\mu$ is significant but not sufficient for claiming new physics. There is however a good prospect that the situation can be clarified in the coming years.

On one hand,  we have seen in this workshop that new $e^+e^-\to {\rm hadrons}$ cross section measurements have been or are being performed by BaBar~\cite{bogdan}. New and precise measurements are expected from CMD3 and SND at VEPP-2000 in Novosibirsk, BES3 in Beijing~\cite{eidelman} and KLOE-2 at DA$\Phi$NE in Frascati~\cite{passer}. These data may help us to resolve a few open issues in the current $e^+e^-$ data and the comparison between the $e^+e^-$ and $\tau$ data, in particular in the $\pi^+\pi^-$ and $\pi^+\pi^-2\pi^0$ channels. At the moment, the $\pi^+\pi^-$ discrepancy between BABAR and KLOE in some of the energy ranges prevents us from achieving a better precision in the data combination. 
Lattice calculations are not yet competitive with the dispersion approach with data but are making significant progress~\cite{bmw13}. The uncertainty of the light-by-light scattering contribution is the next item to improve. Here both measurements of $\gamma^\ast$ physics from BES3 and KLOE-2 and lattice calculations can help.

On the other hand, the uncertainty of the direct measurement (dominated by the statistical precision) is now larger than the total uncertainty of the SM predictions. Two new $g-2$ experiments from Fermilab and J-PARC are being built and an error reduction by a factor of 4 is expected from these experiments in a few years from now~\cite{lancaster}.

The improvements from both sides will either confirm the discrepancy with a much larger significance or point a way to resolve it.

\vspace{5mm}
{\small I am grateful to the fruitful collaboration with my colleagues and friends Michel Davier, Andreas Hoecker, Bogdan Malaescu and Changzheng Yuan.}


\end{document}